\documentclass[review]{elsarticle}

\usepackage{amssymb}
\usepackage[version=3]{mhchem}
\usepackage{graphicx}
\usepackage[disable]{todonotes}
\usepackage{upgreek}
\usepackage{lscape}
\usepackage[english]{babel}

\journal{no journal yet}

\newcommand{\alumina}{\ensuremath{\mbox{Al}_2\mbox{O}_3}}

\newcommand{\aluminaM}{\ensuremath{\mbox{Al}_2\mbox{O}_3\mbox{:C,Mg}}}
\newcommand{\comments}[1]{}
\newcommand{\specialcell}[2][c]{%
  \begin{tabular}[#1]{@{}c@{}}#2\end{tabular}}
\providecommand{\e}[1]{\ensuremath{\times 10^{#1}}}
  
\def\registered{\textsuperscript{\small{{\ooalign{\hfil\raise .00ex\hbox{\scriptsize R}\hfil\crcr\mathhexbox20D}}}}}

\begin{document}

\begin{frontmatter}

\title{Quantitative read-out of \aluminaM-based fluorescent nuclear track detectors using a commercial confocal microscope}

\author[label1]{Greilich, S.}
\author[label1]{Osinga, J.-M.}
\author[label1]{Niklas, M.}
\author[label1]{Lauer, F.M.}
\author[label2]{Bestvater, F.}
\author[label1,label3,label4]{J{\"a}kel, O.}

\address[label1]{Department of Medical Physics in Radiation Oncology, German Cancer Research Center (DKFZ), Im Neuenheimer Feld 280, D-69120 Heidelberg, Germany}
\address[label2]{Light Microscopy Facility, German Cancer Research Center (DKFZ), Im Neuenheimer Feld 280, D-69120 Heidelberg, Germany}
\address[label3]{Heidelberg Ion-Beam Therapy Center, Im Neuenheimer Feld 450, D-69120 Heidelberg, Germany}
\address[label4]{Radiation Oncology, University Hospital of Heidelberg, Im Neuenheimer Feld 400, D-69120 Heidelberg, Germany}

\begin{abstract}
Fluorescent nuclear track detectors (FNTD) show great potential for applications in ion-beam therapy research, such as dosimetry, advanced beam characterization, in-vivo use or as radiobiological assay. A essential feature of FNTDs is their ability to assess the energy loss of single ions yielding for example LET estimations. This article describes the basic characterisations of FNTDs and our read-out system (a Zeiss LSM710 confocal laser scanning microscope) to enable quantative measurements of energy loss.\end{abstract}

\begin{keyword}
\PACS  87.53.Bn \sep 87.55.N- \sep 78.60.-b
\end{keyword}

\end{frontmatter}


\section{Introduction}
Fluorescent nuclear track detectors (FNTD) based on {\aluminaM} single crystals have excellent characteristics for the detection of fast neutrons and swift heavy charge particles (HCPs). By stable radiochromic transformation of fluorescent color centers they allow for the detection and visualization of 3D local dose distribution throughout the detector volume by confocal laser scanning miscroscopy (CLSM) with diffraction-limited resolution. For linear energy transfer (LET) greater than 0.2 keV/$\mu$m, their particle detection efficiency has been reported to be close to 100\,\% \citep{Akselrod2011, Osinga2012a}. For personal and medical dosimetry as well as radiobiological research, they provide a significant superiority to existing technologies such as plastic nuclear track detectors (PNTDs, e.g. CR-39)\cite{Benton1984,Benton2001,Yasuda2006}. They are an attractive candiate for research in ion-beam cancer therapy (IBCT) and have been successfully read-out using a latest-generation commercial CLSM \cite{Greilich2012d}.

Such advanced CLSMs allow the easy adjustment of microscope settings to the variation in sample sensitivity, dose deposited and image quality desired while having to obey constraints such as detector saturation, limited dynamic range and lab time. In this contribution we present a detailed analysis of the readout parameters for a commercial CLSM in order to enable a systematic approach to optimize readout. In addition, our approach allows to perform quantitative measurements. This is a prerequist for IBCT studies beyond binary applications (such as fluence measurements), e.g. single track spectroscopy or microdosimetry.

\section{Materials and Methods}
\subsection{Samples}
We used \aluminaM{} single crystals from Landauer, Inc. having dimensions $4\times8\times0.5\,\mathrm{mm}^3$, polished on one side and irradiated with a variety of radiations (Tab. \ref{tab:samples})\todo{re-check background values with new critical distance routine}. To determine the homogeneity of illumination (flatfield), autofluorescent plastic slides by Chroma Technololy Corp. (www.chroma.com, Part No: 92001) were employed as phantoms.

\begin{center}
\begin{landscape}
\begin{table}
\begin{tabular}{l|r|r|r|r|r|c|r|r|r|r}
	Sample ID & Particle & \specialcell{Energy\\ / (MeV/u)}& \specialcell{$\mathrm{LET}^\mathrm{\alumina}_\infty$\\ / (keV/um)}& Fluence / cm$^{-2}$ & Dose / Gy & Facility & \specialcell{$\eta^\mathrm{adj}_\mathrm{mean}$\\ / MHz} & \specialcell{$\eta^\mathrm{adj}_\mathrm{bkg}$\\ / MHz} & \specialcell{$\eta^\mathrm{adj}_\mathrm{cores}$ \\/ MHz} & \specialcell{SNR\\factor $\hat{S}$} \\
\hline
	ChromaSlide & --- & --- & --- & --- & --- & --- & 0.66 & --- & --- & --- \\
	rm1000-1010 & X-ray & 6 MV & --- & --- & 0 -- 300 & DKFZ & 2.7 -- 240 & --- & --- & --- \\
	rm1011-1015 & $\upgamma$ & Co-60 & --- & --- & 100 -- 10\e3 & DTU Nutech & 198 -- 397 & --- & --- & --- \\
	jmo2009 & $^1\mathrm{H}$ & 220 & 1.4 & 9.8\e{5} & 0.54\e{-3} & HIT & --- & 0.48 & 1.5 & 0.57\\
	sg35041 & $^1\mathrm{H}$ & 3.0 & 35 & 1.3\e6 & 19\e{-3} & MPI-K & --- & 0.56 & 8.3 & 4.4 \\
	jmo2007 & $^{12}\mathrm{C}$ & 91 & 90 & 8.3\e5 & 30\e{-3} & HIT & -- & 0.40 & 5.3 & 2.5 \\
	jmo3901 & $^{12}\mathrm{C}$ & 4.0 & 950 & 2.2\e6 & 0.85 & MPI-K & --- & 0.66 & 17 & 7.7 \\
	rh1 & $^{56}\mathrm{Fe}$ & 9.0 & 7500 & 8.4\e5 & 2.6 & Jyvaeskulae & --- & 0.76 & 200$^\ast$ & 13\\
	rh11 & $^{131}\mathrm{Xe}$ & 5.5 & 61000 & 5.0\e5 & 5.0 & Jyvaeskulae & --- & 0.54 & 52 & 9.0 \\
\end{tabular}
\caption{Samples used in this study. Where applicable, the LET in \alumina{} is given as well as particle fluence and dose (range). The count rates were adjusted for laser power and saturation effects but not for fluorophor effects (see section \ref{sec:formalism}). The SNR factor is defined in Eq. \ref{eq:snrfactor}. $^\ast$The count-rate for rh1 is a rough estimate as it was very close to the APD saturation count-rate. }
\label{tab:samples}
\end{table}
\end{landscape}
\end{center}
 
\subsection{Microscope}
We used the Zeiss LSM710 ConfoCor 3 inverted design CLSM along with its ZEN software (version 2009) and three objective lenses. The read-out procedure (633 nm He-Ne laser, 5\,mW, single APD with a 655 nm longpass emission filter (LP655) is in detailed described in \cite{Greilich2012d}.

\subsection{Microscope control parameters}

The ZEN software allows the user to control (amongst others) the following main parameters for read-out:

\subsection{Software}
\label{sec:software}
Images were processed with ImageJ, a free Java program developed by Wayne Rasband \cite{Rasband1997, Abramoff2004}, version 1.43u and later. All following data processing was done by R \cite{RDevelopmentCoreTeam2010} using a newly developed dedicated package ('FNTD').

\section{Formalism for FNTD signal processing}
\label{sec:formalism}
\subsection{Fluorescence as an estimator of local dose}
Our quantity of interest is the spatial distribution of local dose $D(x,y,z)$. It is related to the concentration $\rho_{\mathrm{F}^{+}_2}$ of transformed $\mathrm{F}^{+}_2 (2 \mathrm{Mg})$ centers by a conversion efficiency function $\hat{f}_\mathrm{conv}$

\begin{equation}
\rho_{\mathrm{F}^{+}_2} = \hat{f}_\mathrm{conv}(\rho_{\mathrm{F}^{2+}_2}, \rho_{\mathrm{F}^{+}_2}, ..., D)\quad.
\end{equation}

$\hat{f}_\mathrm{conv}$ describes the efficiency of radiochromatic transformation processes triggered by HCPs and their secondary particles depending amongst others on the availability of pristine and the concentration of already transformed centers. The local fluorescence yield $Y$ as the ratio of absorbed excitation photon flux $\psi_\mathrm{em}$ and emitted fluorescence $\psi_\mathrm{ex}$ photon flux

\begin{equation}
Y= \frac{\psi_\mathrm{em}}{\psi_\mathrm{ex}} = \hat{f}_\mathrm{QY}(\rho_{\mathrm{F}^{+}_2}, ...) = \hat{f}_\mathrm{QY}(\hat{f}_\mathrm{conv}(D))
\end{equation}

depends again on this concentration (and thus $D$) as well as on the quantum yield $\hat{f}_\mathrm{QY}$ of the radiochromically converted centers. $\hat{f}_\mathrm{QY}$ and $\hat{f}_\mathrm{conv}$ and thus the dose response of the FNTD system are linear over a wide range: $Y = f_\mathrm{QY} \cdot \rho_{\mathrm{F}^{+}_2}$. For higher doses ($D\gtrsim D_0\approx 30\,\mathrm{Gy}$) the system is reported to have an exponential-saturation behaviour \cite{Sykora2010a}\todo{Check if reference is the approriate one}. 

However, $\rho_{\mathrm{F}^{+}_2}(D=0)$ is generally not zero, i.e. there is an \textit{a priori} population of $\mathrm{F}^{+}_2$ centers and FNTDs exhibit a considerable fluorescence even when unirradiated. $\rho_{\mathrm{F}^{+}_2}(D=0)$ can vary from sample to sample (intersample variation) and also within a sample (intrasample variation). We refer to the portion of $Y$ arising from $\rho_{\mathrm{F}^{+}_2}(D=0)$, $Y_\mathrm{bkg}$, as 'background'. $\rho_{\mathrm{F}^{+}_2}(D=0)$ might also affect $\hat{f}_\mathrm{conv}$ but if we neglect this influence we can relate -- assuming an exponential saturation -- the additional fluorescence yield to the local dose by

\begin{equation}
Y_\mathrm{add} = Y_\mathrm{total} - Y_\mathrm{bkg} = f_\mathrm{QY} \cdot f_\mathrm{conv} \cdot (1 - e^{-D/D_0})\quad.
\end{equation}

Using the CLSM we detect a finite volume $V$ and infer $Y$ within $V$ from the fluorescence photon count rate $\eta$ in the detector 

\begin{equation}
\eta_\mathrm{net}=\eta - \eta_\mathrm{DC}= \hat{f}_\mathrm{CLSM}(Y) = f_\mathrm{CLSM} \cdot Y\quad,
\end{equation}

where $\eta_\mathrm{DC}$ is the dark current count rate and $\hat{f}_\mathrm{CLSM}$ summarizes all proportionalities such as incomplete light collection, photon loss due to filters, quantum efficiency of detectors, laser power etc. We assume $\eta_\mathrm{net}$ to be linearily dependend on $Y$ and $f_\mathrm{CLSM}$ to be separable into non-interacting components. These describe amongst others the influence of the laser power, detector saturation effects, the lens used, the pinhole diameter, the lateral position and depth of measurement, the sample orientation, the resolution or other factors:

\begin{align}
f_\mathrm{CLSM} =\ & f_\mathrm{laser}(p) \cdot f_\mathrm{sat}(\eta) \cdot \nonumber \\
& f_\mathrm{lens} \cdot f_\mathrm{pinhole}(d_\mathrm{p}) \cdot f_\mathrm{x,y}(x,y) \cdot f_\mathrm{z}(z) \cdot f_\mathrm{pol}(\Phi) \cdot \nonumber \\
& f_\mathrm{res}(n) \cdot f_\mathrm{other}(...)
\end{align}

In Geiger mode, the LSM710 reports the number of counts per spot position $N$ during the illumination ('dwell') time $\tau$ and a number of repetitive scans $R$. 

$N$ can be related to the count rate $\eta$:

\begin{equation}
N = \hat{f}_\mathrm{readout}(\tau,R,\eta) = f_\mathrm{readout}(\tau,R) \cdot \eta = f_\mathrm{dwell}(\tau) \cdot f_\mathrm{rescan}(R) \cdot \eta
\end{equation}

We assume again linearity and non-interaction. ZEN stores integer values of $N(x,y,z)$ in a $n_x\times n_y \times n_z$ cube in the .lsm-file format which also contains a metainformation on most of the microscope settings.

In summary, we find three categories of proportionalities $f$ for total:
\begin{itemize}
	\item Physics factors ($f_\mathrm{QY}$, $f_\mathrm{conv}$), which are given by the FNTD system in general, the individual sample, the radiation quality etc. and which cannot be influenced.
	\item Fixed or quality changing microscope factors ($f_\mathrm{lens}$, $f_\mathrm{pinhole}$, $ f_\mathrm{x,y}$, ...), which can only partly be influenced by the operator or change the characteristics of an image in a way that makes it impossible to quantitatively compare measurements on a pixel-by-pixel basis. We summarize them as $\widetilde{f}_\mathrm{CLSM}$.
	\item Tweaking factors ($f_\mathrm{laser}(p)$, $f_\mathrm{dwell}(\tau)$, $f_\mathrm{rescan}(R$), ...) that allow the user to adjust the readout to the sensitivity of the individual FNTD, the level of dose deposited, the available read-out time, the desired image quality.
\end{itemize}

We can therefore report measurement results in different levels of comparability. The count rate can be computed by 

\begin{equation}
\eta=\frac{N}{f_\mathrm{dwell}(\tau)\cdot f_\mathrm{rescan}(R)}\quad.
\label{eq:countrate}
\end{equation}

Since the APD detector features a certain dead time, the actual count rate will be higher than the detected if the system is outside its linear counting range close to saturation

\begin{equation}
\eta_\mathrm{actual}=f_\mathrm{sat}(\eta)\quad,
\label{eq:satcountrate}
\end{equation}

and the net countrate by

\begin{equation}
\eta_\mathrm{net}=f_\mathrm{sat}(\eta) - f_\mathrm{sat}(\eta_\mathrm{DC})\quad,
\label{eq:netcountrate}
\end{equation}

where $\eta_\mathrm{DC}$ is obtain from photon counts under representative conditions but laser turned off again via Eq. \ref{eq:countrate}. Eventually, the adjusted net count rate eliminates the influence of the laser power settings

\begin{equation}
\eta^\mathrm{adj}_\mathrm{net}=\frac{\eta_\mathrm{net}}{f(p)}\quad,
\label{eq:adjnetcountrate}
\end{equation}

and can be used to report measurements from different samples under fixed CLSM settings $\widetilde{f}_\mathrm{CLSM}$. 

In addition, many fluorophors can show minor non-linearites at longer dwell time and higher laser powers due to the existence of slowly de-exciting triplet states. These states that get preferebly populated at higher excitation intensities and accumulate during illumination and fluorescence is more efficient for short dwell times and lower laser power \cite{Pawley2006, Borlinghaus2006}. If this is the case, an additional function $f_\mathrm{fluorophor}(p, \tau)$ has to be considered

\begin{equation}
\eta^\mathrm{adj}_\mathrm{net}=\frac{\eta_\mathrm{net}}{f(p)}\cdot f_\mathrm{fluorophor}(p, \tau)\quad.
\label{eq:adjnetcountrate_with_fp}
\end{equation}

Also, color center saturation effects can occur for example due to depletion of the ground state. They were observed for excitation power in the mW range\todo{MA: reference?}. Center saturation could be seamlessly introduced as an addition factor in the presented formalism but since excitation power at the sample location for common CLSMs is usually much smaller we do not take it into account here.

Finally, the local dose $D$ can be assessed by

\begin{eqnarray}
D & = & -D_0\cdot ln \left(1-\frac{Y_\mathrm{add}}{f_\mathrm{QY}\cdot f_\mathrm{conv}}\right)\\
& = & -D_0\cdot ln \left(1-\frac{\widetilde{f}_\mathrm{CLSM}}{f_\mathrm{QY}\cdot f_\mathrm{conv}} \cdot\left(\eta^\mathrm{adj}_\mathrm{net}(\mathrm{total})-\eta^\mathrm{adj}_\mathrm{net}(\mathrm{bkg})\right)\right) \\
& \approx & \frac{D_0}{f_\mathrm{QY}\cdot f_\mathrm{conv}/\widetilde{f}_\mathrm{CLSM}} \cdot\left(\eta^\mathrm{adj}_\mathrm{net}(\mathrm{total})-\eta^\mathrm{adj}_\mathrm{net}(\mathrm{bkg})\right)
\label{eq:dose_from_countrate}
\end{eqnarray}

where the latter applies in the linear part of the dose response curve.

\section{Experiments and Results}
We undertook a series of successive experiements intended to quantify the proportionalities $f$ discussed in section \ref{sec:formalism}. 

\subsection{Detection and scanning}

\subsubsection*{Dependence of count-rate on laser-power $f_\mathrm{laser}(p)$}
\begin{figure}[htbp]
	\centering
		\includegraphics[width = 0.40\textwidth]{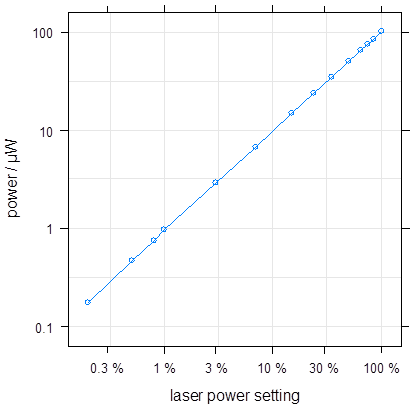}
		\includegraphics[width = 0.45\textwidth]{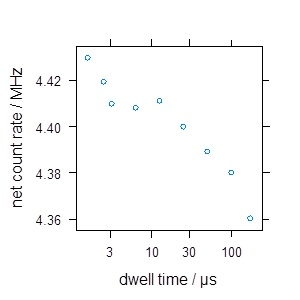}
	\caption{\textit{Laser power measured as function of the microscope setting (left), and net count-rate as function of dwell-time setting (right).}}
	\label{fig:laserpower_clean}
\end{figure}

The laser power at the sample position was measured using a power-meter as a function of the laser power setting $p$ within the ZEN software. The readings exhibited a highly linear relation (exponent: $1.018\pm0.002$, Fig. \ref{fig:laserpower_clean}, left), so we can assume

\begin{equation}
f_\mathrm{laser}(p) = p\quad.
\label{eq:laserpower}
\end{equation}

%
%
%
%

\subsubsection*{Dependence of count-rate on dwell-time $f_\mathrm{dwell}(\tau)$}
Again using reflected light, the dependence of count-rate on dwell-time was assessed for ten dwell times from 1.27\,$\upmu$s to 177\,$\upmu$s. Only a very minor trend towards longer dwell-time was found (Fig. \ref{fig:laserpower_clean}, right) so that we consider the count-rate independent from dwell time and can assume for the proportionality in Eq. \ref{eq:countrate}

\begin{equation}
f_\mathrm{dwell}(\tau) = \tau\quad.
\end{equation}

\subsubsection*{Dependence on resolution and the number of rescans $f_\mathrm{res}(n)$, $f_\mathrm{rescan}(R)$}
A section of sample sg35041 was scanned using settings for the number of spot positions from $n=128$ to $n=6144$ for a fixed image size $l=135\,\upmu$m. The read-out parameters were $p=100\%$, $\tau=177\,\upmu$s and $R=16$ for the smallest $n$ to $\tau=16.8\,\upmu$s and $R=1$ for the highest $n$. The same section of sg35041 was read out with a series of rescans for various settings of $\tau$ and $p$.

\begin{figure}[htbp]
	\centering
		\includegraphics[width = 0.42\textwidth]{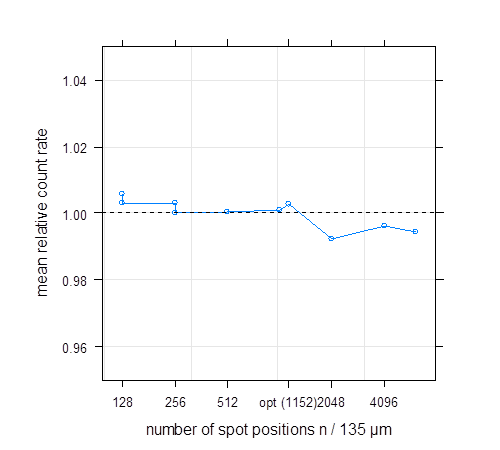}
		\includegraphics[width = 0.57\textwidth]{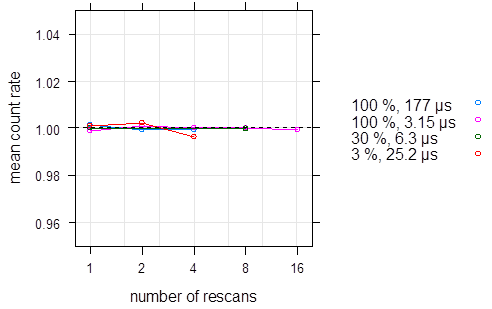}
	\caption{\textit{Dependence of mean count rate $\eta$ from sg35041 with resolution (left) and the number of rescans (right).}}
	\label{fig:resres}
\end{figure}

No significant dependency on $n$ (the 'resolution') was found (Fig. \ref{fig:resres}, left), therefore we conclude that

\begin{equation}
f_\mathrm{res}(n) = 1\quad. 
\end{equation}

Also, no major trend of count rate with the number of rescans could be seen for none of the read-out settings (Fig. \ref{fig:resres}, left), so again we can state that

\begin{equation}
f_\mathrm{rescan}(R) = R\quad. 
\end{equation}

\subsection{Optics}

\begin{figure}[htbp]
	\centering
		\includegraphics[width = 0.45\textwidth]{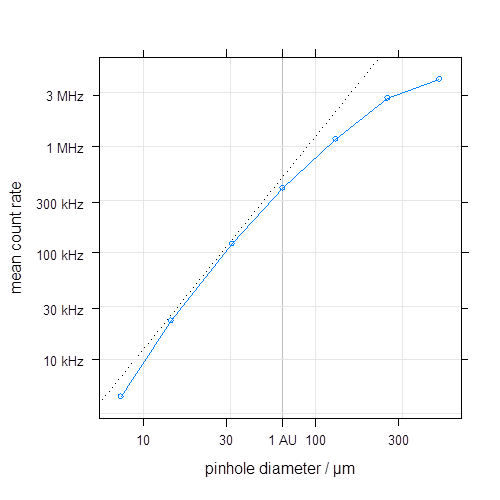}
	\caption{\textit{Increase of count rate with pinhole diameter. The dotted line indicates a quadratic dependency.}}
	\label{fig:mmmpinhole}
\end{figure}

\subsubsection*{Lateral position and depth of measurement $f_\mathrm{x,y}(x,y), f_\mathrm{z}(z)$}

The mean net count-rate from the red autofluorescence plastic slide was measured at approx. 30\,$\upmu$m depth (63x lens, $n = 256$, $l = 135\upmu$m, $\tau = 177\upmu$s, $p=100\,\%$, $R=1$) and as function of depth ('z stack' with $2\,\upmu$m steps from $30\,\upmu$m above to $30\,\upmu$m into the sample with three different pinhole settings, $2\,\upmu$m steps from $30\,\upmu$m to $30\,\upmu$m depth with 1 AU, 63x lens, $n = 512$, $l = 135\upmu$m, $\tau = 2.55\upmu$s, $p=100\,\%$, $R=1$)

\begin{figure}[htbp]
	\centering
		\includegraphics[width = 0.45\textwidth]{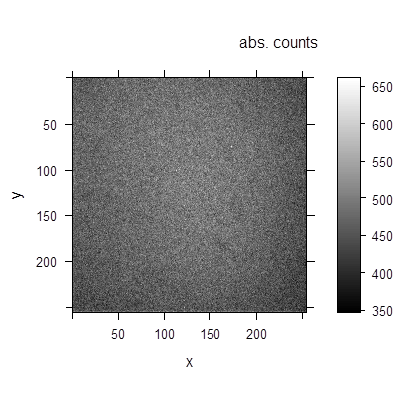}
		\includegraphics[width = 0.45\textwidth]{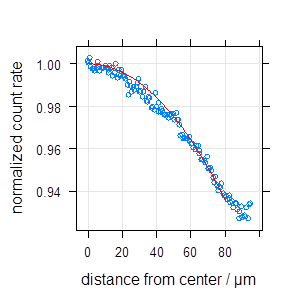}
	\caption{\textit{Image of Chromaslide showing vignetting (left) and normalized count rate as a function of distance from the center (right).}}
\label{fig:vignetting}
\end{figure}

\begin{figure}[htbp]
	\centering
		\includegraphics[width = 0.45\textwidth]{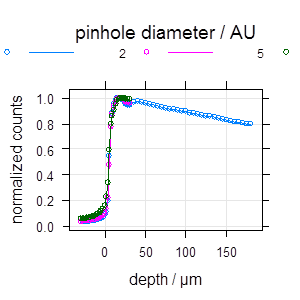}
		\caption{\textit{Normalized count-rate with depth for three different pinhole settings.}}
\label{fig:depth}
\end{figure}

The 63x objective lens shows significant vignetting at a zoom factor of 1 with a signal drop of approx. 6\,\% within 80\,$\upmu$m from the image center (Fig. \ref{fig:vignetting}). The dependency can be well described by

\begin{equation}
f_\mathrm{x,y}= 1-10^{-5}/\upmu\mathrm{ m}\cdot[(x-x_c)^2+(y-y_c)^2]\quad,
\label{eq:depth}
\end{equation} 

where $x_c$ and $y_c$ refer to the image center. In depth, the count-rate shows a sigmoidal transition around the surface of the sample (Fig. \ref{fig:depth}). The width of the transition clearly depends on the pinhole diameter which indicates that it is mainly due to the finite size of the PSF. Beyond approx. 25\,$\upmu$m, $\eta$ drops linearly and can be approximated by

\begin{equation}
 f_\mathrm{z}= 1-10^{-3}/\upmu\mathrm{ m}\cdot [z-z_\mathrm{surface}] 
\label{eq:vignetting}
\end{equation} 

for $z-z_\mathrm{surface}\gtrsim 30\,\upmu\mathrm{m}$.

\subsubsection*{Sample orientation, $f_\mathrm{pol}(\Phi)$}
The mean counts for sample jmo3700 were measured while rotating the sample with respect to the microscope.

The dependency of fluorescence on the angle $\Phi$ between laser polarization direction (parallel to $y$ for the LSM710 used, Fig. \ref{fig:OnMicroscope}) and the sample's c-axis can be expressed as

\begin{equation}
f_\mathrm{pol}(\Phi)=1.00 + 0.58 \cdot \cos (\Phi)
\end{equation}

which means that a drop of $5\,\%$ in signal is only found when exceeding angles of $\pm25\,^{\circ}$.
 
\begin{figure}[htbp]
	\centering
		\includegraphics[width = 0.5\textwidth]{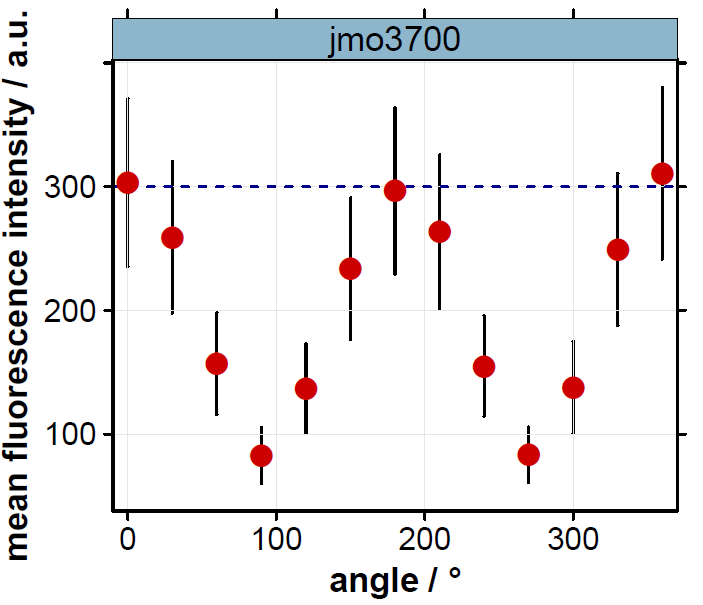}
	\caption{\textit{Mean counts as function of angle between FNTD and laser polarisation axis.}}
\end{figure}

\subsection{Fluorophor effects of dwell time and laser power, $f_\mathrm{fluorophor}(\tau, p)$}

To assess the dependence of fluorescence efficiency, we measured both the red Chroma autofluorescent plastic slide and six FNTD samples (Tab. \ref{tab:samples}) for all combinations of nine laser power ($p=0.2\ldots 100\%$) and ten dwell time ($\tau=1.27\ldots 177 \upmu$s) settings. To minimize systematic effects, the combinations were randomized. All measurements were done using the 63x lens and at $l=135\upmu$m, $n=512$, $d_\mathrm{p}=1\,\mathrm{AU}$. For the plastic slide an additional neutral density filter (T20/R80) was inserted in front of the APD to avoid overload\todo{which is not expected from the actual count rate which did not exceed some MHz}.

\subsubsection*{Chromaslide}
\begin{figure}[htbp]
	\centering
		\includegraphics[width = 0.45\textwidth]{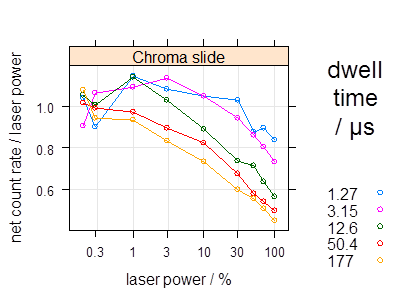}
		\includegraphics[width = 0.45\textwidth]{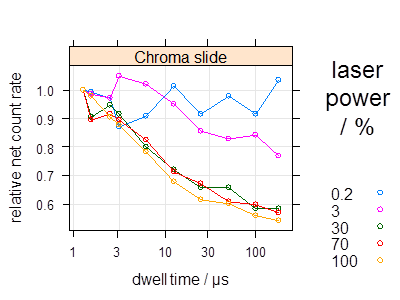}
	\caption{\textit{Deviation of fluoresence (ChromaSlide) from linearity when varying laser power (left) and from constancy when varying dwell time (right).}}
	\label{fig:dw_and_lp_chromaslide}
\end{figure}

A significant drop in effiency up to 50\,\% with increasing laser power and depending on dwell-time settings (Fig. \ref{fig:dw_and_lp_chromaslide}, left) is found. On the other hand, the mean net count-rate is also affected by dwell-time settings (Fig. \ref{fig:dw_and_lp_chromaslide}, right) for higher laser power\todo{Mark: Saturation of center? laser \% not much information but total power, augment discussion}. 
 
\subsubsection*{FNTDs}

\begin{figure}[htbp]
	\centering
		\includegraphics[width = 0.6\textwidth]{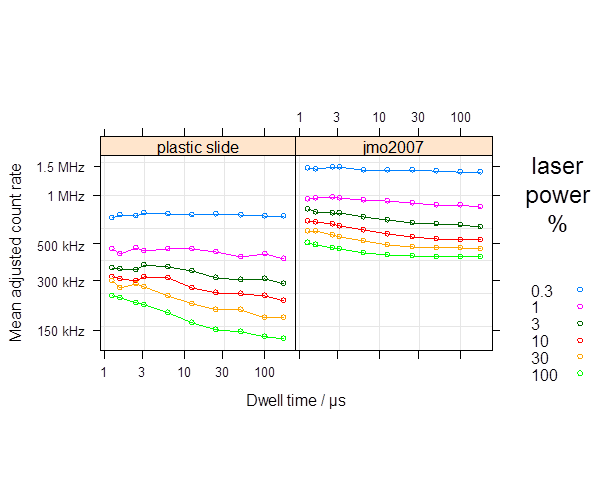}
	\caption{\textit{Mean adjusted count rate for the plastic slide and sample jmo2007 as an example for FNTDs. The data for the plastic slide correspond to those in Fig. \ref{fig:dw_and_lp_chromaslide} but without dark count rate subtraction. $\eta_\mathrm{DC}$ is lower in the case of the plastic slide due to a change in geometry and the additional filtering.}}
	\label{fig:fluorophor_effects}
\end{figure}

\begin{figure}[htbp]
	\centering
		\includegraphics[width = 0.6\textwidth]{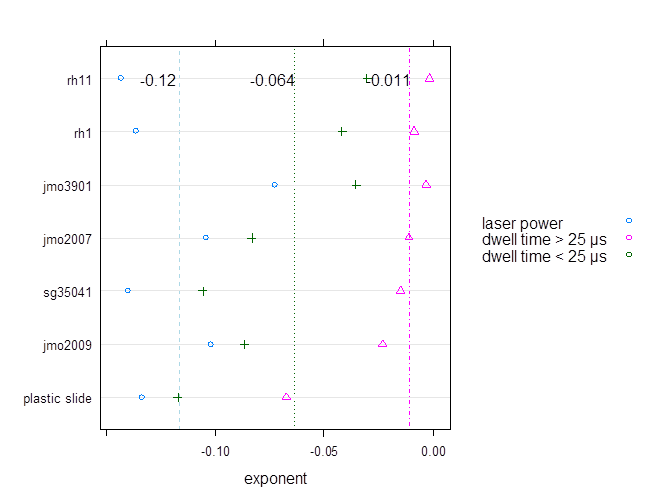}
	\caption{\textit{Exponents for the linear/bi-linear function $f_\mathrm{fluorophor}(p,\tau)$ for the investigated samples and their averages (excluding the plastic slide). No dependency on the radiation used can be found.}}
	\label{fig:fluorophor_exponents}
\end{figure}

FNTDs show similar effects but less pronounced (Fig. \ref{fig:fluorophor_effects}). Effects were quantified using a simple linear regression for $\log p$ and a bilinear fit for $\log \tau$, for which the transition was assumed to be at $\tau=25\,\upmu$s. Interactions between $p$ and $\tau$ were smaller than for the plastic slide and therefore -- also given the uncertainties in the present data -- not considered in the analysis. Also, only data from measurements with $p\ge10$ were included as the dark count rate $\eta_\mathrm{DC}$ constitues a major part of the signal at low laser power. As there is an inherent uncertainty in $\eta_\mathrm{DC}$, a reliable determination of the net signal was not possible. Fig. \ref{fig:fluorophor_exponents} summarizes the exponents that can be used to estimate the effects of changing laser power $p$ and / or dwell time $\tau$:

\begin{equation}
f_\mathrm{fluorophor}(p_1, p_2, \tau_1, \tau_2)=\begin{cases}(p_1/p_2)^{-0.12}\cdot(\tau_1/\tau_2)^{-0.064}, & \mbox{if } \tau <25 \upmu \mbox{s} \\(p_1/p_2)^{-0.12}\cdot(\tau_1/\tau_2)^{-0.011}, & \mbox{if } \tau\ge25 \upmu \mbox{s}\end{cases}
\label{eq:fluorophor_effects}
\end{equation}

\section{Discussion}

\subsection{Detection and scanning}
The microscope proved to be highly linear or constant with respect to many read out settings. Especially, we can rephrase Eq. \ref{eq:countrate} and Eq. \ref{eq:adjnetcountrate} to

\begin{equation}
\eta^\mathrm{adj}_\mathrm{net} = \frac{\eta_\mathrm{net}}{\tau\cdot R\cdot p}
\end{equation}

Also, the user can adapt the resolution $n$ within reasonable limits to their needs without compromising comparability.

The detector dead time, however, is the major deviation from linearity as expressed by $f_\mathrm{sat}$ and yields -6\,\% for 1\,MHz, and already -20\,\% for 4\,MHz. This can be corrected for to some extend but is preferably avoided by using lower laser power and thus high count rates causing saturation of the APDs. We also recommend to refrain from switching the room light on while acquiring images, especially with low laser power as dark counts add up in case of longer dwell times and many rescans. The influence is less problematic when assessing differences such as $Y_\mathrm{add}$ from the same sample since $\eta_\mathrm{DC}$ cancels out in Eq. \ref{eq:dose_from_countrate}. A light-tight cover on the microscope would definitely improve the situation.

\subsection{Optics}
Lenses with smaller numerical aperture deliver higher intensities at equivalent pinhole setting. This is due to the ubiquitous high background in FNTDs and lowers the image contrast significantly (Tab. \ref{tab:dep_on_lens})\todo{My opinion on the goal of readout parameters optimization is not so much maximizing the intensity but maximizing signal to noise ratio of the image where signal is track fluorescent amplitude and noise is 1 s.d. of the background signal. Higher NA and corresponding optimal pinhole size equal to 1 Ary disk should provide the the optimal PSF and assure the best SNR. Reducing the FNTDs background signal also helps a lot.}. Therefore, and due to an apparent tilting of tracks, the 10x lens seems generally not suitable for measurements, while the 63x is to be preferred over the 40x if the smaller size of the imaging field is not of essential importance. In general, the zoom factor should be kept $\le1$, especially for quantative measurements as this minimizes the impact of vignetting effetcs. 

Increasing the pinhole diameter has a similar effect of increasing the available signal on one hand but jeopardizing contrast due to the background. This also explains why epifluorescence microscopy is rarely successful with HCP measurements using FNTDs. Also, changing the pinhole alters the image fundamentally and does not allow to compare track structures on a quantative pixel-by-pixel basis. 

At a measurement depth of approx. 30 $\upmu$m \todo{Mark: 10 um is even enough} the user is far enough from the surface to avoid surface transition effects and at the same time the signal is maximal and its change with depth is smallest\todo{Might be explained by induced spherical abberations as function of depth, could be compensated by cover slip adjustment collar, which we however do not have wit hthe 63x}. When depth profiles (Bragg peak, ranges) are measured, the signal drop with depth should be considered. The sample should also be aligned to the polarization direction of the laser to maximize fluorescence but the signal is relatively robust against small errors in sample angle. The dependency on the angle is in accordance with detailed findings in \cite{Akselrod2003a} (Fig. 2 therein).

\subsection*{Fluorophor effects}
Fluorophor effects are the biggest obstacle against a free variation in microscope read-out parameters in order to optimally adjust to the individual sample. The normalization procedure (Eq. \ref{eq:fluorophor_effects}) should be considered a makeshift only. The recommended approach is to keep dwell time and especially laser power as constant as possible. Improvements in image quality can be achieved by limited variation in $\tau$ (approx. 2\,\% variation was found for a change from $\tau=25\,\upmu$s to $\tau=177\,\upmu$s) or an increased number of rescans. 

\begin{figure}[htbp]
	\centering
		\includegraphics[width = 0.30\textwidth]{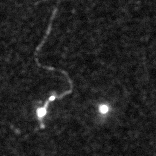}
		\includegraphics[width = 0.30\textwidth]{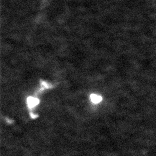}
		\includegraphics[width = 0.30\textwidth]{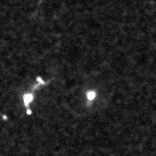}
	\caption{\textit{Read out of same section from a sample irradiated with $^{12}$C at 221\,MeV ($n=156$, $l=18.6\upmu$m). 
	}}
	\label{fig:rescan_test}
\end{figure}

The present microscope, esp. with respect to position stability, its speed and its control software however limits the reasonable use of fast rescanning. In an example (Fig. \ref{fig:rescan_test}), measurements with high laser power and dwell time ($p = 100\,\%$, $\tau = 166.2\,\upmu$s, $R=16$) were compared with series of 80 repeated fast measurements ($p = 10\,\%$, $\tau = 20.8\,\upmu$s, $R=16$). While the mean adjusted count-rates are simliar within 8\,\% for both cases, strong artifacts due to lateral shifts during fast scanning can be seen. Also, the sample tray of microscope lowered during the 25\,min read out time (less then 1 min for the single measurements) and the clearly visible electron track from the first measurement (left panel) disappears. An increase in dwell time should usually be preferred over an increase in the number of rescans.

\section{Conclusion}
This article describes the basic characterisations of FNTDs and our read-out system (a Zeiss LSM710 confocal laser scanning microscope) to enable quantative measurements of energy loss.

\begin{center}
\begin{landscape}
\begin{table}
\begin{tabular}{l|lll}
\hline
1. Lens & \multicolumn{3}{|l}{63x lens, 40x where necessary (, zoom <1)}\\
2. Pinhole & \multicolumn{3}{|l}{1 AU}\\
3. Depth & \multicolumn{3}{|l}{$30\,\upmu$s}\\
4. Quantative read out & \multicolumn{3}{|l}{$p$, $\tau$ as fixed as possible}\\
\hline
\multicolumn{4}{l}{}\\
& \textit{as low as necessary} & \textit{as low as possible} & \textit{as high as possible}\\
\hline
5. Resolution & lab time available & obey Nyquist theorem & resolve relevant structures \\
6. laser power $p$ & \parbox[t]{4 cm}{avoid overload}& \parbox[t]{5 cm}{linear detector response,\\maximum fluorophor efficiency}& \parbox[t]{5 cm}{maximum contrast\\(photon shot noise,\\dark count rate)}\\
7. dwell time $\tau$ & lab time available& maximum fluorophor efficiency & \parbox[t]{5 cm}{maximum contrast\\(photon shot noise)}\\
8. rescans $R$ & lab time available & maximum system stability &\parbox[t]{5 cm}{maximum contrast\\(photon shot noise)}\\
\hline

\end{tabular}
\caption{Recommendations.}
\label{tab:recommendations}
\end{table}
\end{landscape}
\end{center}

\section*{Acknowledgments}
We are deeply indepted to numerous colleagues for their enthusiasm and continuos support for our project: Mark Akselrod and James Bartz from Landauer, Manuela Brom from DKFZ's light microscopy facility for her technical support, Stefan Brons from HIT, Klaus Repnow and Michael K{\"o}nig from MPI-K, Rochus Herrmann from Aarhus University, Jakob Helt-Hansen and Claus E. Andersen from DTU Nutech, Roskilde, and Andrea Schwahofer, Nora H{\"u}nemohr and Roman Martel from DKFZ for their help with FNTD irradiations.

\section*{References}
\bibliography{SteffenBibTex130927}

\begin{thebibliography}{10}

\bibitem{Akselrod2011}
M.~S. Akselrod and G.~J. Sykora.
\newblock Fluorescent nuclear track detector technology - a new way to do
  passive solid state dosimetry.
\newblock {\em Radiation Measurements}, 46:1671--1679, 2011.

\bibitem{Osinga2012a}
J.-M. Osinga, M.S. Akselrod, O.~J{\"a}kel, and S.~Greilich.
\newblock High-accuracy fluence determination in ion beams using fluorescent
  nuclear track detectors.
\newblock {\em submitted to Radiation Measurements}, 2013.

\bibitem{Benton1984}
E.V. Benton.
\newblock Summary of current radiation dosimetry results on manned spacecraft.
\newblock {\em Adv. Space Res.}, 4:153--160, 1984.

\bibitem{Benton2001}
E.R. Benton and E.V. Benton.
\newblock Space radiation dosimetry in low-earth orbit and beyond.
\newblock {\em Nuclear Instruments and Methods B}, 184:255, 2001.

\bibitem{Yasuda2006}
N.~Yasuda, Y.~Uchihori, E.R. Benton, H.~Kitamura, and K.~Fujitaka.
\newblock The intercomparison of cosmic rays with heavy ion beams a nirs
  (icchiban) project.
\newblock {\em Radiation Pr}, 120:414--420, 2006.

\bibitem{Greilich2012d}
S.~Greilich, J.-M. Osinga, F.~Lauer, M.~Niklas, S.~Sellner, J.A. Bartz, M.S.
  Akselrod, and O.~Jäkel.
\newblock Fluorescent nuclear track detectors as a tool for ion-beam
  radiotherapy.
\newblock {\em submitted to Radiation Measurements}, 2013.

\bibitem{Rasband1997}
W.~S. Rasband.
\newblock Imagej, 1997-2011.

\bibitem{Abramoff2004}
M.~D. Abramoff, P.~J. Magelhaes, and S.~J. Ram.
\newblock {Image processing with ImageJ}.
\newblock {\em Biophotonics Int}, 11(7):36--42, 2004.

\bibitem{RDevelopmentCoreTeam2010}
{R Development Core Team}.
\newblock {\em R: A Language and Environment for Statistical Computing}.
\newblock R Foundation for Statistical Computing, Vienna, Austria, 2010.
\newblock {ISBN} 3-900051-07-0.

\bibitem{Sykora2010a}
Garrett~Jeff Sykora and Mark~S. Akselrod.
\newblock Novel fluorescent nuclear track detector technology for mixed
  neutron-gamma fields.
\newblock {\em Radiation Measurements}, 45:594--598, 2010.

\bibitem{Pawley2006}
J.~Pawley.
\newblock {\em Handbook of Biological Confocal Microscopy}.
\newblock Springer, 2006.

\bibitem{Borlinghaus2006}
Rolf~T. Borlinghaus.
\newblock Mrt letter: High speed scanning has the potential to increase
  fluorescence yield and to reduce photobleaching.
\newblock {\em Microscopy Research and Technique}, 69:689--692, 2006.

\bibitem{Akselrod2003a}
Mark~S. Akselrod, Anne~E. Akselrod, Sergei~S. Orlov, Subrata Sanyal, and
  Thomas~H. Underwood.
\newblock Fluorescent aluminum oxide crystals for volumetric optical data
  storage and imaging applications.
\newblock {\em Journal of Fluorescence}, 13:503--511, 2003.

\end{thebibliography}

\newpage
\todototoc
\listoftodos

\end{document}